\documentclass[aps,prb,preprint,showpacs,floatfix,showkeys]{revtex4}
\usepackage{bm}
\usepackage{amssymb}
\usepackage{graphicx}

\begin{document}

\title{Can nano-particle melt
           below the melting temperature of its free surface partner?}

\author{Xiao-Hong Sui}

\affiliation{State Key Laboratory of Theoretical Physics,
         Institute of Theoretical Physics, Chinese Academy of Sciences,
         Beijing 100190, China }

\author{Zongguo Wang}

\affiliation{Computer Network Information Center,
        Chinese Academy of Sciences, Beijing 100190, China}

\author{Kai Kang}

\affiliation{Science and Technology on Surface Physics and Chemistry Laboratory, P.O. Box 718-35, Mianyang 621907, Sichuan, China}

\author{Shaojing Qin}

\affiliation{State Key Laboratory of Theoretical Physics,
         Institute of Theoretical Physics, Chinese Academy of Sciences,
         Beijing 100190, China }

\author{Chuilin Wang}

\affiliation{China Center of Advanced Science and Technology,
            P. O.  Box 8730, Beijing 100080, China}

\date{\today}

\begin{abstract}
The phonon thermal contribution to the melting temperature of
nanoparticles is inspected.
The discrete summation of phonon states and its corresponding
integration form as an approximation for a nanoparticle
or for a bulk system have been analyzed.
The discrete phonon energy levels of pure size effect
and the wave-vector shifts of boundary conditions are
investigated in detail. Unlike in macroscopic thermodynamics,
the integration volume of zero-mode of phonon for a nanoparticle
is not zero,  and it plays an important role in pure size effect
and boundary condition effect. We find that a nanoparticle will have
a rising melting temperature due to purely finite size effect;
a lower melting temperature bound exists for a nanoparticle
in various environments, and the melting temperature of a nanoparticle
with free boundary condition reaches this lower bound.
We suggest an easy procedure to estimation the melting temperature,
in which the zero-mode contribution will be excluded,
and only several bulk quantities will be used as input.
We would like to emphasize that the quantum effect of discrete energy levels in
nanoparticles, which was not present in early thermodynamics studies on finite size
corrections to melting temperature in small systems,
should be included in future researches.
\end{abstract}

\maketitle

\section{INTRODUCTION}
\label{sec:intro}

The analysis of melting temperature might initially seem to be an
area since prehistoric times, but upon reflection it is clear that
this topic is still active for this new material age.
Despite the great strides forward that have been made in nano
technology, the demand for understanding and prediction on
properties of nano-materials continues.
Can nanoparticles melt at a temperature below the one their
freestanding partners melt?  Such a question is raised to
powder ink for the new ``3D printing'' technology.
It is a general question on how much melting temperature
can be changed by surface modification or environment
variation for a nano particle, whether the melting temperature
of a freestanding nanoparticle can be depressed even further
by surface modification, and what the lower bound of the temperature
decreasing is in these changes on boundary conditions of
the nano particle. This work is intended to study physics
of the size dependent melting temperature and the lower limit
of the change on melting temperature. It is an absorbing subject
for decades since it leads to new materials with applications
\cite{STjong:2004,MZhang:2000,MZhao:2004,A-4,RBerry:1988,BK-0}.

Researchers were called upon to present thermodynamics analysis
\cite{PPawlow:1909,CWronski:1967,A-6} for nanoparticles'
melting temperatures \cite{MTakagi:1954,T-1}.
The nanoparticle melting temperature $T_{mn}$ and the bulk melting
temperature $T_{mb}$ have a linear relation \cite{PPawlow:1909,N-2}:
\begin{equation}
T_{mn}/T_{mb}=1-A/D,  \label{eq:linear}
\end{equation}
where $D$ is the nanoparticle diameter
and $A$ is a constant different for different materials.
To understand this linear relation and to find out the constant $A$,
several others analyzed the surface-phonon instability
\cite{KK:2009,C-1,C-2,C-3}; some others
modeled the nanoparticle as a liquid drop \cite{KNanda:2002,KNanda:2006},
or the racks of chemical bonds \cite{WQi:2004,A-8}.
These investigations started from a rational scenario of the melting,
but not yet lead to a converged expression for constant $A$
in simple physical quantities.

In the following sections, we use the Lindemann melting
criterion \cite{KK:2009,FLindemann:1910,A-2}
to calculate the melting temperature of a nanoparticle with different
boundary conditions. We study the constant $A$ with the bulk
physical quantities and the shift of the phonon momentum
on boundary of nanoparticles.
A careful inspection shows that the constant $A$
is determined by the depletion of the zero-mode volume
of phonon, and the additional volume from  phonon wave-vectors with some
zero components.
We suggest a simple procedure to estimate the size dependent decreasing
of melting temperature of nano-particles in certain environment,
in which the constant $A$ is estimated
with the sound velocity and several other physical quantities.

\section{MELTING TEMPERATURE}

We can give the melting temperature of a three-dimensional lattice
by the Lindemann criterion \cite{FLindemann:1910}.
When the ratio of $u$, the square root of the mean square of
atom thermal displacement, to $a$ the lattice constant reaches the
Lindemann critical value $L_c$ at a temperature,
the material melts. The corresponding temperature is the melting
temperature $T_m$.  There are several variations \cite{RBerry:1988}
for the Lindemann criterion and some of criticism on the lacking
of a liquid phase picture behind this criterion
\cite{ALawson:2000,N-29}.  In this work,
we use the original criterion to make the estimation simple:
\begin{eqnarray}
L_c&=&\frac{u(T_m)}{a} . \label{eq:ldm}
\end{eqnarray}

We give the derivation of $u(T)$ in harmonic approximation next.
Some details will be introduced in order to give a better background
for discussions in later sections.
We consider a lattice specified by a set of the vectors $\bm{R_i}$
that locate points,  one atom at each lattice point.
The displacement of an atom from its equilibrium position $\bm{R}_i$
is denoted by $\bm{u}_i$.  Within the harmonic approximation,
the Hamiltonian of the system is
\begin{eqnarray}
H&=&T+\Phi \nonumber \\
&\approx&\frac{M}{2}\sum_{i\alpha}\dot{u}_{i\alpha}^2+\Phi_0
+\frac{1}{2}\sum_{ij\alpha\beta}
\Phi_{\alpha\beta}(\bm{R_i},\bm{R_j})u_{i\alpha}u_{j\beta},
\label{eq:ham}
\end{eqnarray}
where $i=1, 2, \ldots, N$. $N$ is the number of atoms of the
nanoparticle. $\alpha=x, y, z$ and $u_{i\alpha}$ is the $\alpha$th
component of the displacement. $M$ is the mass of the atom.
$\Phi_0$ is the constant potential energy when all the atoms
are in their equilibrium positions, and
$\Phi_{\alpha\beta}(\bm{R_i},\bm{R_j})
=(\partial^2 \Phi/\partial u_{i\alpha}\partial u_{j\beta} )_0$.

For a lattice with translational symmetry, the general solution
can be represented by canonical coordinates $ Q_{k\sigma} $ ---
the vibration modes:
\begin{eqnarray}
u_{i\alpha}(t)=\sqrt{\frac{1}{NM}}\sum_{k\sigma}
  Q_{k\sigma}
e_{ k \sigma \alpha }
\text{e}^{\text{i}[\bm{k}\cdot\bm{R_i}- \omega_{\sigma}(\bm{k}) t]}.
\label{eq:u_Q}
\end{eqnarray}
$\omega_{\sigma}(\bm{k})$ and $ \bm{e}_{ k \sigma } $
are vibration frequency and polarization of vibration mode $ \bm{k}$,
respectively. The Hamiltonian in canonical coordinates is
\begin{eqnarray}
H=\frac{1}{2}\sum_{k\sigma}[P_{k\sigma}^*P_{k\sigma}
+\omega_{\sigma}^2(\bm{k}) Q_{k\sigma}^*Q_{k\sigma}],
\end{eqnarray}
where $P_{k\sigma}$ are the canonical momentums. When quantized
in quantum physics, these vibration modes are called also phonon modes.
The function $\hbar\omega_{\sigma}(\bm{k})$ is phonon dispersion,
and $ \bm{k}$ phonon wave-vector.

The atomic mean-squared thermal vibration amplitude is\cite{calaway}
\begin{eqnarray}
\langle u_{i\alpha}^2\rangle=\sum_{k\sigma}\frac{\hbar\,
e_{ k \sigma \alpha }^2} {NM\omega_{\sigma}(\bm{k})}
[\frac{1}{e^\frac{\hbar\omega_{\sigma}(\bm{k})}{k_B T}-1}
            +\frac{1}{2}],
\end{eqnarray}
where $\langle\,\rangle$ means grand canonical ensemble average.
The square root of the mean square of atom thermal displacement
is given :
\begin{eqnarray}
u(T)&=&\sqrt{\frac{1}{N}
\sum_{i\alpha}\langle u_{i\alpha}^2\rangle} \nonumber \\
&=&\sqrt{\frac{\hbar}{2NM}\sum_{k\sigma}
      \frac{1}{\omega_{\sigma}(\bm{k})
             \tanh[\frac{\hbar\omega_{\sigma}(\bm{k})}{2k_B T}]}}.
\label{eq:uT}
\end{eqnarray}
Melting temperature $T_m$ is obtained by solving Eq.~(\ref{eq:ldm}).

Of course, for a real lattice with a complex cell, one can
calculate $u(T)$ by the exact phonon modes, then estimates
the melting temperature $T_m$ by Eq.~(\ref{eq:ldm}).
Since low energy acoustic phonons dominate
the thermal fluctuation for melting \cite{KK:2009},
one reasonable analysis for different lattices can be given.
We will return to discussion on this point when we show some example
analysis for simple lattices.

\begin{figure}
\includegraphics[width=8cm]{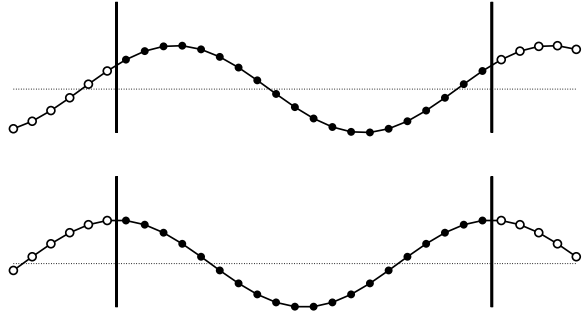}
\caption{
An illustration of phonon modes and boundary conditions
in one-dimension.
The vertical black lines indicate where the boundary is.
The black dotted particle atoms are positioned inside the boundary.
The open circled imaginary atoms are positioned outside the boundary.
The typical atom displacement patterns are plotted
for periodic boundary condition (PBC) on the top
and for free boundary condition (FBC) at the bottom.
}
\label{fig:bcs}
\end{figure}

It is well known that nanoparticles usually possess a lower melting
point than the bulk counterparts for their large surface-volume ratio
\cite{NN-nm}.
More dominated surface effects on melting transition are expected
in nanoparticles \cite{N-20,N-21}, and there are studies on
surface premelting nanoparticles \cite{D-1,C-2,C-4}.
We will investigate the decreasing of melting temperature and the
lower limit of it in present study.
We use boundary conditions to account for various environments.
In next section, we analyze simple ideal periodic boundary
condition (PBC) and free boundary condition (FBC)
for simple cubic nanoparticles.
The ideal boundary conditions are illustrated in Fig.~\ref{fig:bcs} and
defined in following sections.  Melting temperature rising due to
pure size-effect will be uncovered.
The limit of the melting temperature decreasing will be also given.

\section{PBC finite size nanoparticle}

PBC is illustrated in Fig.~\ref{fig:bcs}.
A nanoparticle of a number dimension $L$ with a simple cubic lattice and PBC
has $N=L^3$ atoms in total. It is of size $L$ and has $L$ atoms in each
$\alpha$-direction. The polarization direction can be set
to the coordinate direction. The wave-vector $\bm{k}$ of the phonon
mode $\text{e}^{\text{i}\bm{k}\cdot\bm{R_i}-\text{i}\omega t}$ is given by:
\begin{equation}
(k_x,k_y,k_z) = \frac{2\pi}{La} (n_x,n_y,n_z),
\label{eq:kpbc}
\end{equation}
where the integer $n_\alpha$ runs from $-L/2+1$ to $L/2$, and
$a$ is the lattice constant.
The square root of the mean square of atom thermal displacement
is then given by Eq.~(\ref{eq:uT}).

The summation in Eq.~(\ref{eq:uT}) excludes the $k=0$ mode, since
this zero-mode corresponds to the global moving of the whole nanoparticle
in the three-directions in three dimensional space.
Significant consequences can follow when we exclude zero-mode obviously
in equations for finite size nanoparticles.
With this observation, we calculate the melting temperature $T_{mn}$
for size $L$ nanoparticle by solving Eq.~(\ref{eq:ldm}):
\begin{eqnarray}
&L_c^2&= \frac{u^2}{a^2} \nonumber \\
     &=& \frac{3\hbar a }{16 \pi^3 M}
     \int_{-\pi/a}^{\pi/a} d^3 k \frac{1}{\omega(\bm{k})
     \tanh[\frac{\hbar\omega(\bm{k})}{2k_B T_{mn}}]}
\label{eq:tmpbc}
     \\
     & - & \frac{3\hbar a }{16 \pi^3 M}
     \int_{-\pi/{La}}^{\pi/{La}} d^3 k
          \frac{1}{\omega(\bm{k})
          \tanh[\frac{\hbar\omega(\bm{k})}{2k_B T_{mn}}]}
      , \nonumber
\end{eqnarray}
where ${\omega(\bm{k})}={\omega_{\sigma}(\bm{k})}$ is used when
summing up the three polarized vibration directions in each $\bm{k}$
mode.  The exact magnitude of $L_c$ is not required in our qualitative
estimation of melting temperature.

This transformation for point to box integration is accurate up to
$2\pi/L a$, up to the order of $1/L$ to recover Eq.~(\ref{eq:linear}).
Each mode, $(n_x,n_y,n_z)$, in discrete summation in Eq.~(\ref{eq:uT})
is transformed into a box integration
$ (\frac{La}{2\pi})^3 \int_{-\pi/{La}}^{\pi/{La}} dk_x dk_y dk_z $
centered at point $(k_x,k_y,k_z) = \frac{2\pi}{La}(n_x,n_y,n_z)$.
Stacking all these boxes together in order of their center points
$\bm{k}$, we have the integration.
On each $\alpha$-direction the discrete point
$k_\alpha=2 n_\alpha \pi/La$ runs from $n_\alpha=-L/2+1$ to $L/2$.
Then the integration interval is $[-\pi/a,\pi/a]$ for each integration
variable $k_\alpha$.  The first term in Eq.~(\ref{eq:tmpbc}) is
a full first Brillouin zone(BZ) integration, and the $k=0$ box is included.
The second term is the $k=0$ box, but it is a box to be taken away.
So the $k=0$ mode is excluded.  This $k=0$ integration box is named
as the zero-mode volume in this study.

For the bulk system, melting temperature $T_{mb}$ is the solution
of the same equation but for $(\pi/L a) \to 0$, which turns the
second term into zero in the above integration.
The deduction volume of zero-mode is zero, and only one term is
left in the equation to determine $T_{mb}$:
\begin{equation}
L_c^2 = \frac{3\hbar a }{16 \pi^3 M}
     \int_{-\pi/a}^{\pi/a} d^3 k \frac{1}{\omega(\bm{k})
     \tanh[\frac{\hbar\omega(\bm{k})}{2k_B T_{mb}}]}
      .
\label{eq:tmbulk}
\end{equation}

The zero-mode deduction part in Eq.~(\ref{eq:tmpbc}) in
small $1/L$ limit is approximately
$ \frac{15.35}{L} \frac{3 k_B T_{mn}}{ 8 \pi^3 M v^2}$,
up to the order of $1/L$.
Here we used the relation of the dispersion of acoustic phonon with
the acoustic velocity $v$ in small wave-vector limit:
$\omega(\bm{k})=v k$.  A simple analysis on
Eq.~(\ref{eq:tmpbc}) concludes the melting temperature $T_{mn}$
rising when PBC nanoparticle size $L$ decreases.

This conclusion holds beyond the simple lattice model used
in above analysis.
Taking the phonon modes of any lattice, the mode summation in
Eq.~(\ref{eq:uT}) excludes the $k=0$ zero-mode, which results
in a zero-mode cavity in first BZ integration.  The relation
$\omega(\bm{k})=v k$ holds not only for simple
cubic lattice, but for general situation and for anisotropic
materials in a more complex form.  By the simple lattice model
study in above, we show in the same time that these two steps
are sufficient to conclude an increasing size dependent
melting temperature for nanoparticles subjecting PBC.

This conclusion is contrary to overly repeated words on how the
melting temperature decreases when the nanoparticle size decreases.
Let us briefly picture the microscopic physics here.
As the size $L$ decreases,
the missed volume of zero-mode $(2\pi / La)^3$ increases,
fewer low energy modes contribute to thermal fluctuations
of atoms,
and therefore a higher temperature is required to move atoms
around and to melt the nanoparticle.  PBC is an idea simple boundary
condition but rarely available in experiments.
For PBC the phonon sees no scattering on nanoparticle surface,
it has only the effect from the discreteness of its spectrum.
Without the effect of nanoparticle surface, the increase of
size-dependent melting point is a purely finite size effect.
We provide additional boundary effect with another simple
boundary condition in next section.

\section{FBC finite size nanoparticle}

Let us change the boundary condition of the PBC nano particle
in previous section into free boundary condition (FBC).
The wave-vector $\bm{k}$ for this boundary condition is
\begin{equation}
(k_x,k_y,k_z) = \frac{\pi}{La} (n_x,n_y,n_z), \label{eq:kfree}
\end{equation}
with the integer $n_\alpha$ runs from $0$ to $L-1$.
These wave-vectors are the result of the mode scattering on the
particle surface.
FBC is illustrated in Fig.~\ref{fig:bcs}.  In the illustration,
we show that the boundary atom will experience no external force
from outside of the boundary,
since the displacement of the atom on the other side of the boundary
is the same.
The mode expansion for atom displacement in Eq.~(\ref{eq:u_Q}) for
PBC is changed to the following for FBC:
\begin{eqnarray}
u_{i\alpha}(t)={\textstyle \sqrt{\frac{8}{NM}}}
   & \sum_{k\sigma} & Q_{k\sigma}e_{k\sigma\alpha}
           \text{e}^{-\text{i} \omega_{\sigma}(\bm{k}) t} \cdot \\
   & & \prod_{\alpha'}\cos[k_{\alpha'} (R_{i\alpha'}-a/2)],
\nonumber
\end{eqnarray}
where on each direction we label $R_{i\alpha}$ by numbers
from $1$ to $L$.

Each mode, $(n_x,n_y,n_z)$, in discrete summation
is transformed into a box integration
$ (\frac{La}{\pi})^3 \int_{-\pi/{2La}}^{\pi/{2La}} dk_x dk_y dk_z $
centered at point $(k_x,k_y,k_z)$ $=$ $\frac{\pi}{La}(n_x,n_y,n_z)$.
Equation of melting temperature for FBC is:
\begin{eqnarray}
&L_c^2&=  \frac{3\hbar a }{2 \pi^3 M}
     \int_{-\frac{\pi}{2La}}^{\frac{\pi}{a}-\frac{\pi}{2La}}
            d^3 k \frac{1}{\omega(\bm{k})
            \tanh[\frac{\hbar\omega(\bm{k})}{2k_B T_{mn}}]}
\label{eq:tmfree}
\\
      & - &  \frac{3\hbar a}{2 \pi^3 M}
     \int_{-\frac{\pi}{2La}}^{\frac{\pi}{2La}} d^3 k
          \frac{1}{\omega(\bm{k})
          \tanh[\frac{\hbar\omega(\bm{k})}{2k_B T_{mn}}]}. \nonumber
\end{eqnarray}
The second term removes the $k=0$ box included in the first term.
The zero-mode volume removing in the second term is the same
as PBC case.
However, FBC has additional volume in low energy phonon integration:
$ 8 \int_{\frac{-\pi}{2La}}^{\frac{\pi}{a}-\frac{\pi}{2La}} d^3 k $ will recover
the full first BZ integration and integrations on additional volumes.

The first term in above equation integrates on a volume bigger
than the full first BZ volume.
In above equation the lower bound for the first integration is
neither $-\pi/a$ nor $0$, but a point ${\frac{-\pi}{2La}}$
shifted below zero.
On each $\alpha$-direction the discrete point
$k_\alpha= n_\alpha \pi/La$ runs from $n_\alpha=0$ to $L-1$ for FBC.
Then the integration interval is $[-\pi/2La,\pi/a-\pi/2La]$ for each
integration variable $k_\alpha$.
If $\pi/2La$ for FBC is zero in the first integration in above,
the first term will be equal to an integration over first BZ, and be equal
to the first term for PBC.
We see additional integration volumes for FBC.
These additional integration volumes come from discrete modes
with one or two $n_\alpha$ being zero,
come from the transformation of the sum of discrete points
to the integration for the stack of boxes.

One step further in this simple analysis,
the additional volume will give nonzero terms of order ${\pi/2La}$.
Since the above FBC equation returns to Eq.~(\ref{eq:tmbulk}) for
bulk material in $\pi/2La \to 0$,
in the same microscopic physics pictured in PBC section,
more low energy phonon contribute to atom displacement
in finite size FBC nanoparticle than in bulk material.
At a lower temperature, the thermal phonon population is big
enough, the critical fluctuation magnitude of atom displacement
has been reached, and nanoparticle melts.  So the melting
temperature decreases as a finite size effect.

The melting temperature is bounded below for a particle with
certain fixed size, bounded below by the melting temperature
for FBC.
A model with boundary is good for embedded and deposited
nano particles.  Surface melting and reconstruction are also
modeled as certain boundary.  We analysis the limit of the
shift of the low energy phonon mode towards the zero wave-vector
for different boundary conditions.
Let us arrange the wave-vector $k_\alpha = \frac{2\pi n_\alpha}{La}$
into two different groups: the first group for the integer $n_\alpha$ 
running from $0$ to $L/2-1$, with mode expansion 
$\cos[k_{\alpha} (R_{i\alpha}-(L+1)a/2)]$; the second group for
for the $n_\alpha$ runs from $1$ to $L/2$ with mode expansion
$\sin[k_{\alpha} (R_{i\alpha}-(L+1)a/2)]$.
Inside the first group, wave-vector $k_\alpha$ in each direction
starts from $0$, and the low energy phonon density
can not be increased anymore.  It is not possible to lower the melting
temperature by a boundary condition shifting these wave-vectors upward.
Inside the second group, we can shift the $k_\alpha$ down
towards zero by appropriate boundary conditions.  In scattering
theory, the range of the shift of $k_\alpha$ for low energy phonon in
the second group is $[- \frac{\pi}{La},0]$.  When the shift is zero,
it is for PBC, and the shift $- \frac{\pi}{La}$ is for FBC.
So the low energy phonon modes shift the most for FBC,
the low energy phonon modes density increased most for FBC,
and the melting temperature decreases the most for FBC.
Therefore the melting temperature for size $L$ nanoparticle
can not be below the one for FBC.
This lower bound set by the ideal FBC is close to the melting
temperature for freestanding nano particles in many experiments
in real situation.

The discussion on lower limit of melting temperature
is not applicable when the size $L$ is too small, or $1/L$ is too big.
The analysis in integration functions suggests a smooth scaling of
melting temperature with nanoparticle size with different
boundary conditions.  This is true for large nanoparticles
\cite{N-2,MTakagi:1954,N-4}.
The analysis is qualitatively right if the low energy phonon
dispersion is almost kept while the high energy phonon energy shifted
\cite{KK:2009,HFrase:1998,SCalvo:2005}.
When surface atoms are too many or $1/L$ increases to a certain value
\cite{NN-nm},
we enter the clusters' region. In cluster region the reconstruction
of surface moves the atoms on the nanoparticle surface out of the model boundary, leaves
only a few atoms inside.  The perturbation analysis with a new boundary
condition will not help much.  At the size scale of clusters,
usually atoms are positioned far from the lattice of the bulk
material, and the number of atoms subjects to magic numbers,
and properties fluctuate along with size \cite{N-8}.
The best investigating method is the Ab initio computation
case by case \cite{TLi:2000}.
The limit of the melting temperature depression for clusters
of certain fixed size is a difficult open question.

\section{Qualitative estimation and prediction methods}

When an almost linear result from an experiment is available,
a fitting to the points of $1/L$ and $T_{mn}/T_{mb}$ pairs will give
the slope $A$ in Eq.~(\ref{eq:linear}) for size dependent melting.
A fitting with one more term quadratic in Eq.~(\ref{eq:linear})
can reduce the error of fitting to 1 K \cite{T-3}.
Any physical understanding can suggest a math expression for $A$.
Comparing the expression value and the data fitting value,
we find more physics about the nanoparticle than the experimental data.
An expression is even more powerful.
In interesting nanoparticle designs \cite{A-9},
an expression full of physics will suggest certain bulk
materials be the parent materials, certain kinds of coating
be used to adjust the unfavorable environment.

Thermodynamics of small systems \cite{BK-2} is well understood
in 50 years ago.  Thermodynamics related expressions for constant
$A$ have not converged to physically equivalent ones yet.
For any analytical study captured the main physics behind
the finite size melting phenomena, the size and the surface,
its expression does agree with other similar ones reasonably
good \cite{A-5,MZhao:2004}.
However, zero-mode is the
key for phase transition in statistical physics.
For PBC there is no effect from boundary condition.
The discreteness of phonon spectrum is a pure size effect,
which results in a small but finite zero-mode volume,
and it raises the melting temperature of a nanoparticle with
a boundary condition close to PBC.
Any analytical study failed to display this increase in
melting temperature for PBC like boundary conditions
is not correct on quantum effect of finite size nanoparticle.
Up to now zero-mode volume has not been exactly excluded
in studies\cite{A-1}.
We suggest the exclusion of the zero-mode volume,
a reflection of the discrete levels in final expression,
and the physical equivalence of estimation expressions
be considered more by researchers in the future.

From the phonon zero-mode we inspected in
previous sections, comes a simple procedure to estimate the
melting temperature for almost freestanding nanoparticles.
High quality phonon dispersion can be
calculated \cite{BK-3,A-3} for any particular nano material.
A simple estimation for the melting temperature is not the method
we point out just after Eq.~(\ref{eq:uT}) in Section II.  With that
much effort, one will obtain a rather accurate prediction depending
on details of the dispersion.
Instead, we suggest a simple estimation procedure which uses
$ \omega(\bm{k})= v k $
for all modes in first BZ.
A universal computer program can be coded.
A melting temperature $T_{mn}$ can be generated by the program
in seconds when bulk quantities are input.
This can be a try before any kind of accurate
simulation or calculation started for a nanoparticle design.

We give an example of this procedure for a parent
material of simple cubic lattice and melting temperature $T_{mb}$.
A constant given by bulk properties will be used:
$C = \frac{\hbar v }{2 a k_B}$.
For a nanoparticle of size $L$,
the equation to estimate melting temperature $T_{np}$ for PBC
is given by Eq.~(\ref{eq:tmpbc}) and Eq.~(\ref{eq:tmbulk}):
\begin{eqnarray}
     & T_{mb} & \int_{-\pi}^{\pi} \frac{d^3 k}{k}
\frac{C/T_{mb}}{ \tanh[C k / T_{mb}] }
\\
     &=&
     T_{np} \left[
    \int_{-\pi}^{\pi}
     \frac{\frac{C}{T_{np}} d^3 k}{k \tanh \frac{C k}{T_{np}} }
  - \frac{1}{L} \int_{-\pi}^{\pi}
 \frac{\frac{C}{L T_{np}} d^3 k}{ k \tanh[\frac{C}{ L T_{np}} k ] }
\right]
\nonumber
\\
     &\approx&
     T_{np} \left[
    \int_{-\pi}^{\pi} \frac{d^3 k}{k}
     \frac{\frac{C}{T_{np}}}{ \tanh \frac{C k}{T_{np}} }
  - \frac{15.35}{L}
\right] .
\nonumber
\end{eqnarray}
The term in the form of $x/\tanh(x k)$
is prepared for numerical integration.
A similar equation can be obtained for melting
temperature $T_{nf}$ for FBC.
In this way, we need lattice constant $a$ and sound speed $v$ of the bulk
material. We do not need to calculate the dispersion
$ \omega(\bm{k}) $ for the nanoparticle in this procedure.
When a sound speed can not be found on data bank of bulk materials,
the Young's modulus $Y$ and density $\rho$ can be easily found and
used: $v=\sqrt{Y/\rho}$.  An almost freestanding nanoparticle will
have its melting temperature in between $T_{nf}$ and $T_{np}$.
Its melting temperature will decrease from above to close to $T_{nf}$
if its boundary condition is close to FBC.

\begin{figure}
\includegraphics[width=8cm]{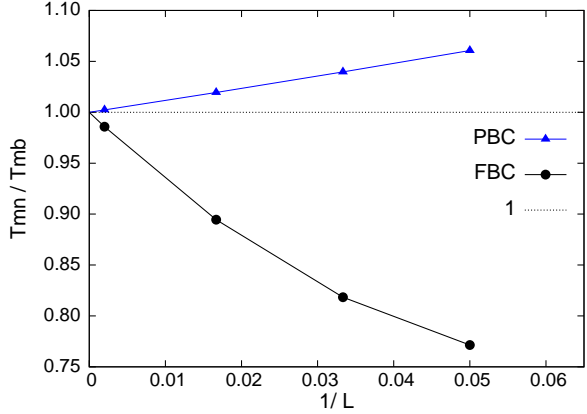}
\caption{(Color online)
The size dependent melting temperature for nanoparticles with PBC and FBC.
The lines and points in the figure are $T_{mn}/T_{mb}$ for PBC and FBC.
A $T_{mn}/T_{mb}=1$ guiding line, and guiding lines simply connecting
the points are presented.  Fittings for the points are
 $1+1.14/L+1.44/L^2$ for PBC, and $1-7.21/L+52.78/L^2$ for FBC.
}
\label{fig:tnptnf}
\end{figure}

At the end of this study, we support our analysis on zero-mode volume
by a numerical calculation\ \cite{KK:2009} on the size-dependent
melting of PBC and FBC.
The particles are of simple cubic lattice, with nearest-neighbor
and next-nearest-neighbor interaction in Hamiltonian Eq.~(\ref{eq:ham}).
We numerically calculate the phonon dispersion $ \omega(\bm{k}) $
for
this simple model\cite{KK:2009}, and obtain the melting temperature
by Eq.~(\ref{eq:ldm}) and (\ref{eq:uT}) displayed in Section II.
In Fig.~\ref{fig:tnptnf} we plot the calculated $T_{mn}$
for nanoparticles with the two boundary conditions and
four different sizes: $L=20,30,60,500$. It should be reminded that $L$ is
the size of a nanoparticle. $L$ is an alias for $D$ in Eq.~(\ref{eq:linear}).

It is shown in Fig.~\ref{fig:tnptnf} that
under PBC the melting temperature increases with the decrease of size,
while under FBC, it is the contrary.
One of the important physical points of this work is successfully
tested by the numerical calculation:
zero-mode volume depletion is important to melting of nano
particles, just as the role of the zero-mode
on phase transition in statistical physics.

\section{CONCLUSIONS}

We investigated in this paper the melting temperatures for nanoparticles
with ideal boundary conditions: periodic and free.  These examples
indicate the scenario of melting: the boundary condition effect
can be separated from pure size effect.
Pure finite size effect, without any boundary to shift the
phonon wave-vectors,
can be accounted by periodic boundary conditions and raises
melting temperature of a nanoparticle as its size decreases.
Boundary conditions play the second important role in melting
temperature design.
We find that the missing of phonon contribution from zero-mode
volume will raise melting temperature, and additional phonon
contribution from wave-vector with some zero components
will depress melting temperature.
The melting temperature for the free boundary condition gives
the lower bound for the melting temperature of a nano particle
in all kinds of environment.

\section*{Acknowledgements}
The authors thank Sen Zhou and visitors to KITPC for discussions.
This work was supported by NNSF1121403 of China.

\end{document}